\font\big=cmbx10 at 12 truept
\def\cl{\centerline} 
\def\vfb{\bar\varphi}


\font\twelverm=cmr10 scaled 1200    \font\twelvei=cmmi10 scaled 1200
\font\twelvesy=cmsy10 scaled 1200   \font\twelveex=cmex10 scaled 1200
\font\twelvebf=cmbx10 scaled 1200   \font\twelvesl=cmsl10 scaled 1200

\font\twelvett=cmtt10 scaled 1200   \font\twelveit=cmti10 scaled 1200
 
\skewchar\twelvei='177   \skewchar\twelvesy='60
 
 
\def\twelvepoint{\normalbaselineskip=12.4pt
  \abovedisplayskip 12.4pt plus 3pt minus 9pt
  \belowdisplayskip 12.4pt plus 3pt minus 9pt
  \abovedisplayshortskip 0pt plus 3pt
  \belowdisplayshortskip 7.2pt plus 3pt minus 4pt
  \smallskipamount=3.6pt plus1.2pt minus1.2pt
  \medskipamount=7.2pt plus2.4pt minus2.4pt
  \bigskipamount=14.4pt plus4.8pt minus4.8pt
  \def\rm{\fam0\twelverm}          \def\it{\fam\itfam\twelveit}%
  \def\sl{\fam\slfam\twelvesl}     \def\bf{\fam\bffam\twelvebf}%
  \def\mit{\fam 1}                 \def\cal{\fam 2}%
  \def\tt{\twelvett}
  \textfont0=\twelverm   \scriptfont0=\tenrm   \scriptscriptfont0=\sevenrm
  \textfont1=\twelvei    \scriptfont1=\teni    \scriptscriptfont1=\seveni
  \textfont2=\twelvesy   \scriptfont2=\tensy   \scriptscriptfont2=\sevensy
  \textfont3=\twelveex   \scriptfont3=\twelveex  \scriptscriptfont3=\twelveex
  \textfont\itfam=\twelveit
  \textfont\slfam=\twelvesl
  \textfont\bffam=\twelvebf \scriptfont\bffam=\tenbf
  \scriptscriptfont\bffam=\sevenbf
  \normalbaselines\rm}
 
 
\def\tenpoint{\normalbaselineskip=12pt
  \abovedisplayskip 12pt plus 3pt minus 9pt
  \belowdisplayskip 12pt plus 3pt minus 9pt
  \abovedisplayshortskip 0pt plus 3pt
  \belowdisplayshortskip 7pt plus 3pt minus 4pt
  \smallskipamount=3pt plus1pt minus1pt
  \medskipamount=6pt plus2pt minus2pt
  \bigskipamount=12pt plus4pt minus4pt
  \def\rm{\fam0\tenrm}          \def\it{\fam\itfam\tenit}%
  \def\sl{\fam\slfam\tensl}     \def\bf{\fam\bffam\tenbf}%
  \def\smc{\tensmc}             \def\mit{\fam 1}%
  \def\cal{\fam 2}%
  \textfont0=\tenrm   \scriptfont0=\sevenrm   \scriptscriptfont0=\fiverm
  \textfont1=\teni    \scriptfont1=\seveni    \scriptscriptfont1=\fivei
  \textfont2=\tensy   \scriptfont2=\sevensy   \scriptscriptfont2=\fivesy
  \textfont3=\tenex   \scriptfont3=\tenex     \scriptscriptfont3=\tenex
  \textfont\itfam=\tenit
  \textfont\slfam=\tensl
  \textfont\bffam=\tenbf \scriptfont\bffam=\sevenbf
  \scriptscriptfont\bffam=\fivebf
  \normalbaselines\rm}
 

\def\beginparmode{\endmode
  \begingroup \def\endmode{\par\endgroup}}
\let\endmode=\par
{\obeylines\gdef\
{}}
\def\singlespace{\baselineskip=\normalbaselineskip}
\def\oneandathirdspace{\baselineskip=\normalbaselineskip
  \multiply\baselineskip by 4 \divide\baselineskip by 3}

\def\doublespace{\baselineskip=\normalbaselineskip \multiply\baselineskip by 2}

\newcount\firstpageno
\firstpageno=2
\footline={\ifnum\pageno<\firstpageno{\hfil}\else{\hfil\twelverm\folio\hfil}\fi}
\let\rawfootnote=\footnote              
\def\footnote#1#2{{\rm\singlespace\parindent=0pt\rawfootnote{#1}{#2}}}
\def\raggedcenter{\leftskip=4em plus 12em \rightskip=\leftskip
  \parindent=0pt \parfillskip=0pt \spaceskip=.3333em \xspaceskip=.5em
  \pretolerance=9999 \tolerance=9999
  \hyphenpenalty=9999 \exhyphenpenalty=9999 }
\def\dateline{\rightline{\ifcase\month\or
  January\or February\or March\or April\or May\or June\or
  July\or August\or September\or October\or November\or December\fi
  \space\number\year}}
\def\received{\vskip 3pt plus 0.2fill
 \centerline{\sl (Received\space\ifcase\month\or
  January\or February\or March\or April\or May\or June\or
  July\or August\or September\or October\or November\or December\fi
  \qquad, \number\year)}}
 
 
\hsize=6.5truein
\hoffset=.1truein
\vsize=8.9truein
\voffset=.05truein
\parskip=\medskipamount
\twelvepoint            
\doublespace            
\overfullrule=0pt       
 
\def\preprintno#1{
 \rightline{\rm #1}}    

\def\head#1{                    
  \filbreak\vskip 0.5truein     
  {\immediate\write16{#1}
   \raggedcenter \uppercase{#1}\par}
   \nobreak\vskip 0.25truein\nobreak}
 
\def\references                 
  {\head{References}            
   \beginparmode
   \frenchspacing \parindent=0pt \leftskip=1truecm
   \parskip=8pt plus 3pt \everypar{\hangindent=\parindent}}
 
\def\frac#1#2{{\textstyle{#1 \over #2}}}
\def\square{\kern1pt\vbox{\hrule height 1.2pt\hbox{\vrule width 1.2pt\hskip 3pt
   \vbox{\vskip 6pt}\hskip 3pt\vrule width 0.6pt}\hrule height 0.6pt}\kern1pt}
\newcount\pagenumber
\newcount\sectionnumber
\newcount\appendixnumber
\newcount\equationnumber

\newcount\citationnumber
\global\citationnumber=1

\def\ifundefined#1{\expandafter\ifx\csname#1\endcsname\relax}
\def\cite#1{\ifundefined{#1} {\bf ?.?}\message{#1 not yet defined,}
\else \csname#1\endcsname \fi}

\def\docref#1{\ifundefined{#1} {\bf ?.?}\message{#1 not yet defined,}
\else \csname#1\endcsname \fi}

\def\article{
\def\eqlabel##1{\edef##1{\sectionlabel.\the\equationnumber}}
\def\seclabel##1{\edef##1{\sectionlabel}}    
\def\citelabel##1{\edef##1{\the\citationnumber}{\global\advance\citationnumber by1}}
}

\def\appendixlabel{\ifcase\appendixnumber\or A\or B\or C\or D\or E\or
F\or G\or H\or I\or J\or K\or L\or M\or N\or O\or P\or Q\or R\or S\or
T\or U\or V\or W\or X\or Y\or Z\fi}

\def\sectionlabel{\ifnum\appendixnumber>0 \appendixlabel
\else\the\sectionnumber\fi}

\def\beginsection #1
 {{\global\subsecnumber=1\global\appendixnumber=0\global\advance\sectionnumber by1}\equationnumber=1
\par\vskip 0.8\baselineskip plus 0.8\baselineskip
 minus 0.8\baselineskip 
\noindent $\S$ {\bf \sectionlabel. #1}
\par\penalty 10000\vskip 0.6\baselineskip plus 0.8\baselineskip 
minus 0.6\baselineskip \noindent}

\newcount\subsecnumber
\global\subsecnumber=1

\def\subsec #1 {\bf\par\vskip8truept  minus 8truept
\noindent \ifnum\appendixnumber=0 $\S\S\;$\else\fi
$\bf\sectionlabel.\the\subsecnumber$ #1
\global\advance\subsecnumber by1
\rm\par\penalty 10000\vskip6truept  minus 6truept\noindent}

\def\beginappendix #1
{{\global\subsecnumber=1\global\advance\appendixnumber by1}\equationnumber=1\par
\vskip 0.8\baselineskip plus 0.8\baselineskip
 minus 0.8\baselineskip 
\noindent
{\bf Appendix \appendixlabel . #1}
\par\vskip 0.8\baselineskip plus 0.8\baselineskip
 minus 0.8\baselineskip 
\noindent}

\def\no{\eqno({\rm\sectionlabel} 
.\the\equationnumber){\global\advance\equationnumber by1}}

\def\ref #1{{\bf [#1]}}

\article
\def\i{\infty}  
\def\L{\Lambda} 
\def\l{\lambda} 
\def\o{\over}   

         \def\p{\partial}
\def\G{\Gamma}   \def\k{\kappa}  
\def\ra{\rightarrow}           \def\ft{\vf^2}                 \def\cf{{\cal F}}
\def\e{\varepsilon}    
\def\gf{\gamma_{\vf}}              \def\vfb{\bar\vf}
  
\def\vf{\varphi}
 
\def\gft{\gamma_{\vf^2}}  
  
\def\frac#1#2{{{#1}\over{#2}}}
\def\s{\scriptstyle}     \def\ss{\scriptscriptstyle}

\def\gl{\gamma_{\l}}

\def\io{\hbox{$\bigcirc\kern-2.3pt{\bf\cdot}$\kern 2.3pt}}
\def\bub{\hbox{${\bf\cdot}\kern-5pt\bigcirc\kern-5pt{\bf\cdot}$}}     
\def\itro{\hbox{${\bf\cdot}\kern-5pt\bigcirc\kern-5pt{\bf :}$}}
\def\bfour{\hbox{${\bf :}\kern-5pt\bigcirc\kern-5pt{\bf :}$}}

\def\a{\alpha}

\def\de{d_{\ss eff}}

\def\ef{_{\ss eff}}

\def\Gnl{\G^{(N,L)}}
\def\R{\rlap I\mkern3mu{\rm R}}

\oneandathirdspace
 at 17.3 truept

\vskip -0.4truein
\preprintno{ICN-UNAM-94-09}
\preprintno{DIAS-STP-95-25}
\dateline
\cl{\bf \big Dimensional Crossover in the Large N Limit}
\bigskip 
\cl{\bf\big Denjoe O'Connor}
\cl{Dublin Institute for Advanced Studies} 
\cl{10 Burlington Road,
Dublin 4, Ireland}
\bigskip
\cl{\bf\big C. R. Stephens}
\cl{I.C.N., U.N.A.M,}
\cl{Circuito Exterior, A. Postal 70-543}
\cl{M\'exico D.F. 04510, M\'exico} 
\bigskip
\cl{\bf\big A. J. Bray}
\cl{Department of Theoretical Physics,}
\cl{The University, Manchester M13 9PL, U.K.}
\bigskip
\bigskip \bigskip 
\noindent{\bf Abstract:}\ \ We consider dimensional crossover for an
$O(N)$ Landau-Ginzburg-Wilson model on a $d$-dimensional film geometry of
thickness $L$ in the large $N$-limit. We calculate the full universal
crossover scaling forms for the free energy and the equation of state.  We
compare the results obtained using ``environmentally friendly''
renormalization with those found using a direct, non-renormalization group
approach. A set of effective critical exponents are calculated and scaling
laws for these exponents are shown to hold exactly, thereby yielding
non-trivial relations between the various thermodynamic scaling functions.
\vfil \eject

\beginsection{Introduction}
Crossover behavior --- the interpolation between qualitatively different
effective degrees of freedom of a system as a function of scale --- is an
ubiquitous phenomenon in nature. Calculation of scaling functions
associated with crossover behavior is, generally speaking, much more
difficult than the calculation of critical exponents, the latter being
calculable in an approximation scheme suitable for the asymptotic region
around one critical point. 

An important non-trivial and experimentally accessible example is seen in
the context of dimensional crossover. As far as the fluctuations in a
system are concerned there is a marked difference
between an ``environment'' consisting of infinite three dimensional space
and a three dimensional box of ``size'' $L$. By implementing a
renormalization programme which is explicitly dependent on the relevant 
environmental parameters one obtains a globally defined RG which 
incorporates several fixed points at once and with which one 
can calculate the crossover scaling functions perturbatively using 
one uniform approximation scheme. Recently such a
program which allows for the treatment of a wide class of crossovers has
been developed
\citelabel{\NPhysOrig}\citelabel{\Btcfss}\citelabel{\proc}\citelabel{\fintemp}
\citelabel{\prl}\citelabel{\EnvRG}[\cite{NPhysOrig}-\cite{EnvRG}] 
and dubbed with the epithet --- ``environmentally friendly''
renormalization in recognition of the fact that a crossover is often 
induced by some relevant perturbation which can be associated with the
effects of the ``environment'' on a system. 

The gist of the approach is based on the intuition that a ``good'' coarse
graining will be one that when effected to a length scale comparable to
any set by the environment, will reflect the influence of the
latter by continuously changing as a function of scale the 
type of effective degree of freedom being coarse
grained. However, although the
intuition is grounded in such Kadanoff/Wilson type coarse graining,
such a procedure is notoriously difficult to implement.  Instead the
formalism we adopt is based on the field theoretic RG which represents
the invariance under reparametrization of the couplings of the system, an
idea which goes back to the original formulation of the RG in the '50's. 

Just as there are good and bad coarse grainings so too there are good and bad
reparametrizations. An environmentally friendly reparametrization is one
that tracks the qualitatively changing nature of the effective degrees of
freedom for a crossover system. As has been emphasized previously
[\cite{EnvRG}] a necessary condition for an environmentally friendly RG to
satisfy is that the number of fixed points of the RG, defined globally on
the space of parameters, be isomorphic to the number of points of scale
invariance of the system. Unfortunately, for many crossovers, favorite
forms of field theoretic renormalization, such as minimal subtraction, do
not satisfy this criterion and therefore are of only limited use in
describing crossover behavior. 

Various new results, both formal and perturbative, have been obtained,
predominantly in the context of dimensional crossover both above and below
[\cite{Btcfss}] the critical temperature. The formalism has also been
applied to several other crossover systems [\cite{proc}], and has been
implemented in particle physics [\cite{fintemp}]. Two loop Pad\'e resummed
results have been obtained for the case of an $O(N)$ model on a layered
geometry, and also for a quantal Ising model [\cite{prl},\cite{EnvRG}].
Effective exponents were calculated which asymptotically are in excellent
agreement with known results \citelabel{\baker}[\cite{baker}].
Additionally, where available, there is good agreement between the results
of environmentally friendly renormalization and numerical results from
lattice simulations \citelabel{\binder}[\cite{binder}] and high
temperature series \citelabel{\capefish}[\cite{capefish}] approximations. 

Exactly solvable models provide a useful testing ground for ideas on phase
transitions and quantum field theory. The growth of interest in conformal
field theory in two dimensions, and the growing number of exactly solvable
models in this same dimension, is evidence of the interest in such models.
In this paper we wish to test environmentally friendly renormalization
thoroughly in the context of an exact model.  To date the only truly
exactly solvable models, in the sense that all correlation functions are
calculable, in any dimension, are the Gaussian and the Berlin-Kac spherical model
\citelabel{\BerlinKac}[\cite{BerlinKac}], and variants of these.  However,
all higher vertex functions above two are identically zero for the
Gaussian model and it has no ordered phase, pathologies absent in the
spherical model. Stanley \citelabel{\Stanley}[\cite{Stanley}] established
the equivalence for an infinite lattice of the partition functions of the
$N=\infty$ limit of the $O(N)$ vector sigma model and the spherical model. 
The latter model was discussed from a field theoretic point of view by
Wilson \citelabel{\Wilson}[\cite{Wilson}] whose analysis led to subsequent
developments where the model served as the beginning point of a
perturbative expansion in $1/N$ (see
\citelabel{\BreWadia}[\cite{BreWadia}] for a detailed set of reprints on
this topic).  The original lattice
spherical model was solved for both strictly finite geometries and
geometries exhibiting a dimensional crossover by Barber and Fisher
\citelabel{\BarberFisher}[\cite{BarberFisher}].  (See Rudnick
\citelabel{\Rudnick}[\cite{Rudnick}] for a more recent discussion
of the model in a purely finite geometry).  
While Allen and Pathria \citelabel{\AllenPathria}[\cite{AllenPathria}] 
have recently studied the models two point correlation function.

We choose as our testing ground the limit $N\ra\i$ of an
$O(N)$ Landau-Ginzburg-Wilson model. This model is closely related to the
spherical model and the $O(N)$ sigma model, but it contains an additional
parameter $\l_B$ the $\varphi^4$ coupling, which away from the critical point
governs the crossover to mean field behavior. 
It is the field theoretic formulation that we discuss in the following,
our interest being the model in a film geometry.
The purpose of the present paper
is to present a solution of the Landau-Ginzburg-Wilson field theory
variant of the spherical model from two points of view and 
analyze both the universal and non-universal aspects of the dimensional 
crossover both above and below the critical point. 

The format of the paper is as follows: in section 2 we give a brief
overview of the large-$N$ limit. We then derive, via a
saddlepoint evaluation of the partition function, 
the scaling functions which incorporate the
dimensional crossover and the crossover to mean field theory for the free
energy and equation of state.  We analyze in detail the universal scaling
limit of these functions and compare with known results for the spherical
model. In section 3 we analyze the model with the environmentally friendly
RG approach and demonstrate how the results of section 2 are recovered. 
In section 4 we calculate a set of effective exponents which are scaling
functions that describe the crossovers between $d$-dimensional,
$d-1$-dimensional and mean field fixed points.  In section 5 we show that
the effective exponents satisfy natural analogs of the standard scaling
laws, including hyperscaling. 
Finally we make our conclusions in section 6. 

\beginsection{The Large $N$ limit.}
The  $O(N)$ ``microscopic'' Landau-Ginzburg-Wilson Hamiltonian 
for a $d$ dimensional film geometry of thickness $L$ is given by
\eqlabel{\ham} 
$${\cal H}[{\bf\vf}]=\int_0^L\int 
d^{d}x\left({1\o2}\nabla\varphi^a\nabla\varphi^a+ 
{1\o2}r_{B}(x)\varphi^a\varphi^a 
+{\l_B\o4!}(\varphi^a\varphi^a)^2-H^a(x){\varphi^a}\right)\no$$ 
where $r_B(x)=m_B^2+t_B(x)$. We will restrict our considerations 
to the case of a film which exhibits
critical behavior ($d\ge3$) and make the standard assumptions that 
$\l_B$ is temperature independent and that all the temperature dependence
of the model is contained in the variable $t_B$.

The partition function $Z$ is obtained by performing the path integral
over the order parameter fields, ${\vf^a}$, with the Hamiltonian
(\docref{ham}). The generator of one particle irreducible vertex
functions, $G[\vfb]$, where $\vfb^a$ is the induced magnetization, is the
Legendre transform of $W[H]=-\ln Z$. If
the sources $H^a$ and $t_B$ are taken to be homogeneous then for a 
translationally invariant system $\vfb^a$ is also homogeneous and in the 
direction of $H^a$ and $G[\vfb]=V\G[\vfb]$ where $V$ is the volume. 
It is convenient however to retain the general case for the
moment. The vertex functions $\G^{({\s a_1...a_N})}$ are the objects
of primary interest to us as once these are known all the correlation
functions of the theory can be reconstructed from them.  To avoid over
encumbering the notation we have dropped a conventional subscript $B$,
referring to ``bare'' quantities, from the fields and vertex functions. 

In general for the $O(N)$ model there are two types of modes: Those along
the direction picked out by the field, $H^a$, and those perpendicular to
it. If we choose the direction of the field to be given by the unit vector
$n^a$ then using the two projectors
$$P^{ab}_l=n^an^b \qquad\qquad P^{ab}_t=\delta^{ab}-n^an^b$$
we can decompose a general vertex function into block diagonal form. We
denote a generic vertex function by $\G^{(N)}_{l\dots lt\dots t}$ where
the number of $l$ and $t$ subscripts indicates whether a longitudinal or a
transverse propagator is to be attached to the vertex at the corresponding
point.  When all subscripts are either $l$ or $t$ we will use a single $l$
or $t$, for example $\G^{(N)}_{t\dots t}$ will be abbreviated
$\G^{(N)}_t$. 

Due to the Ward identities of the model it is sufficient to know only the
$\G^{(N)}_t$ as all the other vertex functions can be reconstructed from
these.  Thus for example the equations of state, using the Ward identity
$\G^{(1)}_l=\G^{(2)}_t\vfb$, become \eqlabel{\basiceqnofstate}
$$\G^{(2)}_t\vfb=H\qquad\qquad \G^{(1)}_t=0\no$$ 
Decomposing $\G^{(ab)}$ yields $\G^{(2)}_{l}$, $\G^{(2)}_{t}$ and
$\G^{(2)}_{lt}$.  Ward identities imply that
$$\G^{(2)}_{l}=\G^{(2)}_{t}+{\G^{(4)}_{t}\over3}\vfb^2\qquad
\hbox{and}\qquad\G^{(2)}_{lt}=0\no$$

The large $N$ limit is taken such that $N\l_B$ is held fixed as
$N\ra\infty$. In this setting it is possible to obtain exact expressions
for the vertex functions of the theory.  One can do this either by a
direct resummation of the Feynman diagrams or via a saddle point
approximation. In the latter approach the identity
$$e^{-{\l\o4!}\int\vf^4}=A\int d\psi 
e^{\int({N\o2}\psi^2-({\l N\o12})^{1\o2}\psi\vf^2)}\no$$
where $A$ is a normalization constant, allows one to perform the now Gaussian
$\vf$ integral, which yields the effective Hamiltonian
$${\cal H}_{\ss eff}\equiv{\cal H}_{\ss eff}(\psi,H)
=-\int\left\{{N\over2}\psi^2+{1\over2}{H^2\over m_B^2+t_B+\sqrt{{N\l_B\over3}}\psi}\right\}+{NV\over2}\bigcirc_{\ss \L}
\no$$
We use the diagrammatic notation of [\cite{EnvRG}], where
$$\bigcirc_{\ss\L}={Tr\over V}\ln\left[-\nabla^2+m_B^2+t_B+\sqrt{{N\l_B\over3}}\psi\right]\no$$
and $(-1)^{k-1}/(k-1)!$ times the $k$'th derivative with respect to $t_B$
will be represented by a circle with $k$ dots, the dots representing the
point at which each derivative acts.  The subscript $\L$ represents the 
presence of an ultraviolet cutoff $\L$ that regulates the 
diagram. In the large $N$ limit the integral
over the auxiliary field $\psi$ can be done in a saddle point
approximation, the saddle point condition being
\eqlabel{\stepestdescent}
$${\partial{\cal H}_{\ss eff}\over\partial M^2}=0\no$$
where $M^2=m_B^2+t_B+\psi\sqrt{N\l_B\o3}$. The effective Hamiltonian
${\cal H}_{\ss eff}$ evaluated
on the solution of (\docref{stepestdescent}) then yields the leading large
$N$ behavior of $W[H,M^2]$ and a Legendre transform yields
$\Gamma[\vfb,M^2]+\Gamma_{reg}[t_B,\L]=W+H\vfb$ where 
$$\vfb=-{\partial W\over\partial H}.\no$$ 
which we have split into a
singular part $\Gamma[\vfb,M^2]$, which vanishes at the bulk critical 
point, and a remaining regular part $\Gamma_{reg}[t_B,\L]$. 
The function $\G[\vfb,M^2]$ determines the singular part of the free
energy density to which it is related by $k_BT$. 

For the translationally invariant case with homogeneous $H$ 
the saddle point constraint (\docref{stepestdescent}) 
together with (\docref{basiceqnofstate}) lead to 
\eqlabel{\constone}
$$M^2\vfb=H\no$$ 
and 
\eqlabel{\consttwo}
$$M^2=m_B^2+t_B+{\l_B\over6}\vfb^2+{N\l_B\over6}\io_{\ss\L}\no$$
The former is just the
equation of state and the latter specifies the relationship between temperature
and the transverse correlation length $M^{-1}$.

Using (\docref{consttwo}) the singular part of  
$\Gamma$ can be expressed in the form
\eqlabel{\gfreeen}
$$ \G={3M^4\o2\l_B}+{N\over2}(\bigcirc-M^2\io)\no$$
while the regular part is given by
$$\Gamma_{reg}[t_B,\L]=-{3(m_B^2+t_B)^2\o 2\l_B}+d_d\L^d\no$$
In (\docref{gfreeen}) any cutoff dependence of the diagrams is 
subtracted so that they vanish at the bulk critical point. 
{}From equation (\docref{consttwo}) one finds
\eqlabel{\msquare}
$$\left.{dM^2\over d\vfb}\right\vert_{\ss t_B}={\l_B\over 1+{\l_BN\over6}\bub}{\vfb\over3}\no$$
and 
$$\left.{dM^2\over dt_B}\right\vert_{\ss \vfb}={1\over1+{N\l_B\over6}\bub}\no$$
We have further the vertex functions
\eqlabel{\gzeroone}
$$\G^{(0,1)}={3M^2\over\l_B}\no$$
\eqlabel{\gzerotwoba}
$$\G^{(0,2)}={3\over\l_B}{1\over1+{N\l_B\over6}\bub}\no$$
\eqlabel{\gzerothree}
$$\G^{(0,3)}={N\itro\over {(1+{N\l_B\over 6}\bub)}^2}\no$$
and 
$$\G^{(2)}_{l}=M^2+{dM^2\over d\vfb}\vfb\no$$
where we focus on the singular parts of these vertex functions only.

For later convenience, with $H=0$, we define ${\cal E}=\G^{(0,1)}$ and ${\cal C}=-\G^{(0,2)}$.
When the temperature dependence of  $t_B$ is taken to be 
$t_B=\Lambda^2{(T-T_c)\over T}$, we have ${\cal E}$ and ${\cal C}$  
proportional to the  internal energy and specific heat 
respectively \citelabel{\spheat}[\cite{spheat}].
Given the Ward identities we need only specify the even transverse vertex 
functions $\G^{(N)}_t$ since the others can be obtained from these. Thus 
$$\G^{(2)}_t=M^2\no$$
$$\G^{(4)}_t={\l_B\over1+{\l_BN\over6}\bub}\no$$
$$\G^{(6)}_t={5\over9}{\l_B^3N\itro\over {(1+{\l_BN\over6}\bub)}^3}\no$$

Note that the diagrammatic structure of all expressions is the same
irrespective of whether we are treating the bulk problem, a film geometry,
or the case of a completely finite geometry.  However, the functions
represented by the diagrams differ in the different cases. In the
dimensional crossover problem each diagram depends explicitly on $L$, the
film thickness. For a $d$-dimensional layered geometry ($d<4$) with
periodic boundary conditions in the infinite cutoff, $\Lambda\ra\i$, limit
\eqlabel{\perbub} 
$$\bub(M^2,L)={(d-3)\over2}\sigma_{d-1}
\sum_{n=-\infty}^{\infty}{1\over{[M^2+{({2\pi n\over L})}^2]}^{(5-d)/2}}\no$$
where
$$\sigma_d=-{\G({2-d\over2})\over {(4\pi)}^{d/2}}\no$$
This diagram is well defined for $d<4$ and $M\neq0$ and can be used to specify 
the other diagrams from which it is derived by differentiation. The first of these 
is the tadpole, $\io$, which when required to vanish at the bulk critical 
point is given by
\eqlabel{\taddef}
$$\io(M^2,L)=-\int^{\ss M^2}_0\bub(x,L)dx+{b_d\o L^{d-2}}\no$$
where
\eqlabel{\bddefn}
$$b_d={\G({d-2\over2})\zeta(d-2)\over2{\pi}^{d/2}}.\no$$
The other is the vacuum diagram which in the same limit and again 
required to vanish at the bulk critical point ($M=0$ and $L=\infty$) is 
\eqlabel{\vacdiag}
$$\bigcirc(M^2,L)=-\int^{\ss M^2}_0dx\int^x_0dy\bub(y,L)
+{b_d\o L^{d-2}}M^2-{a_d\o L^d}\no$$
where 
$$a_d={2\Gamma(d/2)\zeta(d)\over\pi^{d/2}}\no$$
is a universal number associated with the critical theory 
defined  by 
$${1\over2}a_d=
L^d\left\{\left.\Gamma\right\vert_{T_c(\infty)}-\left.\Gamma\right\vert_{T_c(L)}\right\}\no$$
 
The critical temperature is that temperature for which $M=0$ and $\vfb=0$.
It is therefore determined by the zero of the right hand side of
(\docref{consttwo}) with $\vfb=0$. Now, since $t_B$ by definition becomes
zero at $T_c(L)$, (\docref{consttwo}) determines $m_B^2$ to be such as to
cancel the last term in (\docref{taddef}) and any cutoff dependence that
would have given the diagram a non-zero value at the bulk critical point.
This of course implies that $m_B^2$ depends on $L$ and can be written as
$m_B^2=r_B-\Delta(L)$ where $\Delta(L)$ vanishes for
$L\rightarrow\infty$ and $r_B$ is $L$ independent. {}From (\docref{taddef})
we see that $\Delta(L)={N\l_B\over 6}{b_d\over L^{d-2}}$. Likewise we
can decompose $t_B$ into an $L$ independent part which would then vanish
at $T_c(\infty)$ and an $L$ dependent part. 

It is not difficult to see
that the corresponding decomposition of $t_B$ yields $t_B(\infty)+\Delta(L)$. 
With the temperature dependence
$$t_B(\infty)=\Lambda^2{(T-T_c(\infty))\over T}\no$$
discussed in [\cite{spheat}], and the fact that
$t_B(\infty)$ vanishes at the bulk critical temperature 
we see that
$$\Lambda^2{T_c(\infty)-T_c(L)\over T_c(L)}=\Delta(L)={N\l_B\over 6}{b_d\over L^{d-2}}\no$$
and so $\Delta(L)$ is proportional to the shift in critical temperature of the film 
from the bulk critical temperature. Since $b_d$ is positive we see
that the film critical temperature is suppressed relative to the bulk one
and scales with the shift exponent $d-2=1/\nu(d)$, $\nu(d)$ being the bulk 
correlation exponent all of which is in agreement with the lattice results of 
Barber and Fisher [\cite{BarberFisher}]. Furthermore
since $b_d$ diverges at $d=3$ we see that for a three dimensional film 
the critical temperature $T_c(L)$ is driven to zero and more careful analysis 
is appropriate.

In terms of the variable $t_B$
(\docref{consttwo}) becomes
\eqlabel{\sphcons}
$$M^2=t_B+{\l_B\over6}\vfb^2+{N\l_B\over6}\io'\no$$
where $\io'=-\int_0^{M^2}dx\bub(x,L)$. 
Normalizing so that $\Gamma$ vanishes at the bulk critical temperature
$T=T_c(\infty)$ with zero external field $H=0$
we have
\eqlabel{\freeenfin}
$$\G={3M^4\over2\l_B}+{N\o2}\left(\bigcirc'-M^2\io'-{a_d\over L^d}\right)\no$$
where $\bigcirc'=-\int^{\ss M^2}_0dx\int^x_0dy\bub(y,L,\i)$. 

Explicitly, 
one finds
$$\G[M,L]={N\over2}\left\{{3M^d\over g}+{{\cal G}(d,z)-a_d\over L^d}\right\}\no$$
where $g=N\l_BM^{d-4}$, $z=LM$, and  
the scaling function ${\cal G}(d,z)$ for periodic boundary conditions is given by
$$\eqalign{
{\cal G}(d,z)
&=z^{d-1}\sigma_{d-1}
\left\{
{d-3\over d-1}-{4\over d-1}\sum_{n=1}^{\infty}\left\{
{\left[1+{({2\pi n\over z})}^2\right]}^{(d-1)\over2} \right.\right.\cr
&\left.\left.\qquad\qquad\qquad
-{\left({2\pi n\over z}\right)}^{(d-1)}
-{(d-1)\over2}{\left[1+{({2\pi n\over z})}^2\right]}^{(d-3)\over2}
\right\}\right\}\cr}\no$$
which makes the small $z$ behavior manifest. 
A convenient integral representation of ${\cal G}$ is  
$$\eqalign{{\cal G}(d,z)=
&{d-2\over d}\sigma_dz^d+ z^{d-1}\left(
{d-3\over d-1}\sigma_{d-1}\right.\cr
&\quad\left.-{4\over{(4\pi)}^{(d-1)/2}\Gamma({d-1\over2})}
\int_0^\infty dk k^{d-4}
(k^2+{d-3\over2})
\ln\left[{1-e^{-kz}\over k}{\sqrt{1+k^2}\over1-e^{-\sqrt{1+k^2}z}}\right]\right)
\cr}$$
For $d=3$ we find the simpler result 
\eqlabel{\threedimgofz}
$${\cal G}(3,z)={z^3\over12\pi}+{1\over2\pi}\int_0^zdy {y^2\over e^y-1}\no$$
with $a_3=\zeta(3)/\pi$.

The transverse correlation length $M^{-1}$ is determined 
in terms of  $t_B$ and $\vfb$ by 
\eqlabel{\constrnt}
$${3M^2\over\l_B}={3t_B\over\l_B}+{1\over2}\vfb^2-{NM^{d-2}\over2}{\cal F}(d,z)\no$$
where  
${\cal F}=-{\io'\over M^{d-2}}$ and for periodic boundary conditions is given explicitly by
\eqlabel{\fscaln}
$$\eqalign{
{\cal F}(d,z)=&{\sigma_{d-1}\over z}
\left\{1+2
\sum_{n=1}^{\infty}\left[
\left(1+{({2\pi n\over z})}^2\right)^{(d-3)/2}-
{\left({2\pi n\over z}\right)}^{(d-3)}
\right]
\right\}\cr
=&\sigma_d+{1\over z}\left(\sigma_{d-1}-{4\over{(4\pi)}^{(d-1)/2}\Gamma({d-3\over2})}
\int_0^\infty dk k^{d-4}\ln\left[{1-e^{-kz}\over k}{\sqrt{1+k^2}\over1-e^{-\sqrt{1+k^2}z}}\right]\right)
\cr}
\no$$
Re-arranging  the constraint equation (\docref{constrnt}) we can express it in terms of 
$\tau={6t_B(\infty)\over N\l_B}$ with $t_B(\infty)$ the $L$ 
independent temperature parameter which vanishes at the bulk critical 
temperature and $\tilde\varphi=\vfb/\sqrt{N}$. 
We then have (\docref{constrnt}) in the form 
\eqlabel{\scalingform}
$$w\equiv(\tau+{b_d\over L^{d-2}}+\tilde\varphi^2)L^{d-2}
={z^{d-2}\over g}+z^{d-2}{\cal F}(d,z)\no$$
and we see that $\tau(L)={6 t_B\over N\l_B}$ is given by 
$$\tau(L)=\tau+\Delta(L) \qquad\hbox{with}\qquad\Delta(L)={b_d\over L^{d-2}}\no$$
If we consider the small $\l_B$ limit we obtain mean field results, while
the universal scaling regime is governed by the limit $\l_B\rightarrow\infty$
in which case the universal form of the free-energy per component  
$\tilde\Gamma$ is given by
\eqlabel{\scalingfn}
$$\tilde\Gamma={{\cal G}(d,z)-a_d\over2L^d}\no$$
The universal form of $M[\tau,\tilde\varphi,L]$ is determined by
\eqlabel{\invscaling}
$$w={\cal Q}^{-1}(z^2)=z^{d-2}{\cal F}(d,z)\no$$ 
In terms of the basic scaling variables 
$$\tilde x= (\tau+{b_d\over L^{1/\nu(d)}})\vert\tilde\varphi\vert^{-1/\beta(d)}
\qquad\hbox{ and}\qquad
\tilde y=L\vert\tilde\varphi\vert^{\nu(d)/\beta(d)}\no$$
with $\nu(d)={1\over d-2}$ and $\beta(d)={1\over2}$ the bulk $d$
dimensional exponents
$$w=(1+\tilde x)\tilde y^{1/\nu(d)}\no$$
Note that $\tilde x$ could equally well be
expressed in the form 
$\tilde x=(b_d+\tau L^{1/\nu(d)})\tilde y^{-1/\nu(d)}$ 
so in general there is a choice of variables in which the scaling
function can be expressed. For the large $N$ limit, irrespective of whether we
consider the universal limit or not, we see that in general only the
combination $\tau+\vert\tilde\varphi\vert^{\ss{1\o\beta}}$ 
plays a r\^ole. This has
significant consequences for the effective exponents to be considered
later, since there is a reduction from two variable scaling functions to
scaling in terms of the single variable
$w=(\vert\tilde\varphi\vert^{\ss{1\o\beta}}+\tau) L^{1/\nu}+b_d$. 
 
Equation (\docref{invscaling}) implies that\footnote{*}{More generally
if we do not take the universal limit from (\docref{scalingform})
we have the more general two variable scaling form
$z^2={\cal Q}(d,v,w)$ where $v=N\l_BL^{d-4}$}
\eqlabel{\scalingconstrt}
$$z^2={\cal Q}(d,w)\no$$
Substitution of (\docref{scalingconstrt}) into (\docref{scalingfn}) yields 
the scaling function ${\cal G}(d,w)$ for $\tilde\G$.
Finally the equation of state is given by
\eqlabel{\univQeqofst}
$${\cal Q}(d,w)\tilde\varphi L^{-1/\beta(d)}=\tilde H\no$$
 where $\tilde H=H/\sqrt{N}$. The asymptotic forms of the 
equation of state then become: For $L\rightarrow\infty$ 
\eqlabel{\infLeqofstate}
$$\sigma_d^{-\gamma(d)}(1+\tilde x)^{\gamma(d)}\tilde\varphi^{\delta(d)} =\tilde H\no$$
and for $L\rightarrow0$  
\eqlabel{\smallLeqofstate}
$${({L\over\sigma_{d'}})}^{\gamma(d')}(1+\tilde x)^{\gamma(d')}\tilde\varphi^{\delta(d')} 
=\tilde H\no$$
where $\delta(d)=(d+2)/(d-2)$ and $d'=d-1$.
Both limiting forms (\docref{infLeqofstate}) and (\docref{smallLeqofstate}) agree
with the usual universal form of the equation of state 
\citelabel{\Joyce}[\cite{Joyce}] aside from the factors of $\sigma_d^{-\gamma(d)}$
and ${({L\over\sigma_{d'}})}^{\gamma(d')}$ which 
could be absorbed into a redefinition of $\tilde\varphi$ and $\tilde H$. 
We choose not to absorb dimension dependent or $L$ dependent 
factors into our variables as we are interested in a problem involving 
two dimensions at once with $L$ the interpolating physical variable. 

Similarly $\tilde\Gamma$ interpolates between 
$$\tilde\Gamma=\rho_d(1+\tilde x)^{2-\alpha(d)}
\tilde\varphi^{(2-\alpha(d))/\beta(d)}
\no$$
for $L\rightarrow\infty$ 
and
$$\tilde\Gamma=L^{1-\alpha(d')}\rho_{d'}(1+\tilde x)^{2-\alpha(d')}
\tilde\varphi^{(2-\alpha(d'))/\beta(d')}+{a_d\over 2 L^d}\no$$ 
for $z\rightarrow0$
where
$$\rho_d={\alpha(d)\over2}\sigma_d^{\alpha(d)-1}
\qquad\qquad\hbox{and }\qquad \alpha(d)={d-4\over d-2}.$$

For $d$ approaching three, we saw that the film critical temperature $T_c(L)$ is 
driven to zero as  $b_d$ diverged with a simple pole at $d=3$. 
However,  $\sigma_{d-1}$ also diverges with a simple pole in this limit and in fact 
the scaling function ${\cal Q}^{-1}(d,z)$ yields a simple pole divergence also. 
If we examine (\docref{taddef}) more carefully we see 
that 
$$\io(M^2,L)=-{M\over 4\pi} -{1\over2\pi L}\ln[1-e^{-z}]\no$$
and so the pole contributions cancel and we are left with 
the universal form of the constraint (\docref{scalingform}) in the form
\eqlabel{\scalingformthreed}
$$(\tau+\tilde\varphi^2)L={z\over 4\pi}+{1\over2\pi}\ln[1-e^{-z}]\no$$
in agreement with [\cite{BarberFisher}], who restricted their considerations
to the  zero field case, $H=0$.  {}From (\docref{scalingformthreed}) we see 
\eqlabel{\threedimq}
$${\cal Q}(3,w)=\left\{2\ln\left[{e^{2\pi w}+\sqrt{e^{4\pi 
w}+4}\over2}\right]\right\}^2\no$$
where we find it convenient to define $w=(\tau+\tilde\varphi^2)L$. 
Then (\docref{threedimq}) together with (\docref{univQeqofst}) specifies 
the universal equation of state.
Similarly (\docref{scalingfn}) with (\docref{threedimgofz}) and (\docref{threedimq}) 
specifies $\tilde\Gamma$.

For $L\rightarrow\infty$ we have $w\rightarrow\infty$ and we recover 
the three dimensional scaling function (\docref{infLeqofstate})  
discussed above, and for  fixed $L$ with 
$\xi_L=M^{-1}\rightarrow\infty$  (\docref{scalingformthreed}) gives
$$w={1\over 2\pi} \ln z\no$$
so in the two dimensional critical regime which is governed by 
$\tau\rightarrow-\infty$ we find
$$\xi_L=L e^{-2\pi(\tau+\tilde\varphi^2)L}.\no$$
The limiting form of the equation of state becomes
$$e^{4\pi(\tau(L)+\tilde\varphi^2)}\tilde\varphi=\tilde H\qquad
\hbox{where}\qquad\tau(L)=\tau-{1\over 2\pi L}\ln L\no$$
in agreement with [\cite{BarberFisher}].

The other special dimension of interest is $d=4$ where $\bub$ has an ultraviolet 
divergence. In this case it is inappropriate to simply send $\l_B\rightarrow\infty$ to 
recover the universal properties. 
For $d=4$, when $\l_B$ is sent to infinity
in such a way as to cancel the divergent contribution from $\bub(4,M,L,\L)$
and render $\G^{(4)}_t$ finite we find the constraint
retains the logarithmic corrections to scaling and becomes
$$ (\tau+{1\over12 L^2}+\tilde\varphi^2)L^2={z^2\over{(4\pi)}^2}
\ln({z^2\over z_0^2})+{z\over4\pi}\left\{1-{2\over\pi}\int_0^\infty dk
\ln[{1-e^{-kz}\over k}{\sqrt{1+k^2}\over1-e^{-\sqrt{1+k^2}z}}]\right\}\no$$
where $z_0=\kappa L$ with $\kappa$ a remnant microscopic scale, such that 
$M\ll\kappa$.
Similarly the free energy scaling function is given by
$$\eqalign{{\cal G}(4,z)=&{z^4\over32\pi^2}(1-\ln{z^2\over z_0^2})
-{1\over3\pi^2}\int_0^\infty dqq^2\left\{
{(q^2+{3\over2}z^2)
\over\sqrt{q^2+z^2}(e^{\sqrt{q^2+z^2}}-1)}-{q\over e^q-1}\right\}
}\no$$
with  $a_4=-1/45$.
For $M\rightarrow0$ with fixed $L$ we recover the three dimensional 
results above,
and for $L\rightarrow\infty$ the constraint becomes
$${(\tau+\tilde\varphi^2)\over\kappa^2}={1\over {(4\pi)}^2}
{M^2\over\kappa^2}\ln({M^2\over\kappa^2})\no$$
while $\tilde\G$ becomes 
$$\tilde\G={M^4\over64\pi^2}(1-\ln{M^2\over\kappa^2}).\no$$

To conclude,  we have found that the universal limit
$\l_B\rightarrow\infty$ of the continuum Landau-Ginzburg-Wilson model
recovers the results of Barber and Fisher for the spherical model
[\cite{BarberFisher}]. Our results further interpolate between the integer
dimensions they considered and we incorporate the presence of a homogeneous
external field $H$ and the crossover to mean field theory. 
We found the scaling functions for the free energy
and equation of state. 

\beginsection{Environmentally Friendly Renormalization} 
The purpose of this section is to use renormalization group techniques to
recover the scaling functions in the large $N$ limit.  We will treat the
problem from the perspective of environmentally friendly renormalization.
As before we assume that the finite system also exhibits critical behavior
and that $3\leq d\leq4$.  We will restrict our considerations to a film
geometry with periodic boundary conditions, the geometry of interest being
$S^1\times\R^{d-1}$. Some of the analysis has appeared in other work
[\cite{EnvRG}], however, for the sake of completeness we will review the
techniques and summarize the main results. 
 
We indicate by $t_B(M)$ that temperature parameter which yields the 
transverse correlation length $\xi_L=M^{-1}$. The renormalization group method
is to change from the original bare parameters to new renormalized ones. 
The renormalized parameters and vertex functions are related to the bare
ones by
$$t(M,\kappa)=Z_{\varphi^2}^{-1}t_B(M)\qquad\qquad 
\qquad\l(\kappa)=Z_\l(\kappa)\l_B
\qquad\qquad\vfb(\kappa)=Z_{\varphi}^{-1/2}\vfb_B$$ 
and 
$$\G^{(N,L)}=Z_{\varphi}^{{N\over2}}(\k)Z_{\varphi^2}^L(\k)\G_B^{(N,L)}
+\delta_{N0}\delta_{Ln}A^{(n)}(\k)\qquad\qquad i=0,1,2\no$$
where $\k$ is an arbitrary renormalization scale.
The renormalized vertex functions then obey the RG equation
\eqlabel{\rgee} 
$$\k{d\o d\k}\Gnl+(L\gft-{N\o2}\gf)\Gnl=\delta_{N0}\delta_{Ln}B^{(n)}\no$$ 
where the Wilson functions in the above are
$\gf={d{\rm ln}Z_{\varphi}\o d{\rm ln}\k}$ and $\gft=-{d{\rm ln}Z_{\varphi^2}\o 
d{\rm ln}\k}$. The final Wilson function is 
$\gl={d{\rm ln}Z_{\l}\o d{\rm ln}\k}$ and is related to the beta function, 
$\beta(\l)$ through the relation $\gl=\beta(\l)/\l$.
Equation (\docref{rgee}) is inhomogeneous for the three 
vertex functions $\G$, $\G^{(0,1)}$ and $\G^{(0,2)}$; which are 
related to the free energy, internal energy and the specific heat 
respectively. By an appropriate set of normalization conditions the Wilson functions
are related to the anomalous dimensions of the operators $\vf$ and $\ft$ respectively. 

The normalization conditions which fix the particular parameterization we
will use to describe physical quantities are,
\eqlabel{\ncone} 
$$ \left.\G_{t}^{ (2)}\right\vert_{NP}=\k^2 \no$$ 
\eqlabel{\nctwo}
$$ \left.{{\p}\o{\p k^2}}\G_{t}^{(2)}\right\vert_{NP}=1 \no$$ 
\eqlabel{\ncthree}
$$ \left.\G_{t}^{(4)}\right\vert_{NP}=\l  \no$$ 
\eqlabel{\ncfour}
$$ \left.\G_{t}^{(2,1)}\right\vert_{NP}=1 \no$$ 
where the subscript $t$, introduced earlier, refers to the 
transverse components of the respective vertex functions, and $NP$ refers to the 
normalization point. 
A simplifying feature in considering the large $N$
limit is that the separate dependence on $\vfb$ and $t_B$ disappears
into a dependence on the transverse correlation length. We choose 
our normalization point to be at zero momentum and transverse correlation 
length, $\k^{-1}$.
Away from the large $N$ limit explicit account must be taken of the fact
that one has a two scale rather
than a one scale problem. The general $O(N)$ case will be dealt with 
elsewhere. 

In the normalization conditions (\docref{ncone}--\docref{ncfour}) the film thickness
is treated as a fixed passive variable. As has been emphasized on 
previous occasions [\cite{NPhysOrig},\cite{EnvRG}] such 
``environmentally friendly'' conditions are essential in order to 
obtain a perturbatively controllable description of the finite size 
crossover. The conditions (\docref{ncone}--\docref{ncfour}) above 
specify the Wilson functions $\gft$, $\gf$ 
and $\gl$, which are explicitly $L$ dependent 
and interpolate between those of a $d$ and $d-1$ dimensional model
as the ratio of film thickness to 
transverse correlation length ranges from infinity to zero. 

The condition (\docref{ncone})
is just the spherical constraint which we analyzed earlier 
and is therefore of a somewhat different 
status to the rest, serving to fix the relationship between the parameter $\k$ 
and the physical variables $t$, $\bar\varphi$ and $\l$. Essentially 
(\docref{nctwo}-\docref{ncfour}) determine the Wilson functions in 
terms of an arbitrary, fiducial transverse correlation length $\k^{-1}$, while 
Eq. (\docref{ncone}) determines the relationship between this 
correlation length and the temperature and magnetization.

In general we saw that it is sufficient to know the properties of 
the transverse vertex functions as the longitudinal ones could be
generated directly from the Ward identities. 
The task now is to exhibit the relationship between the transverse
correlation length, temperature and magnetization and thereby generate the
equation of state for comparison with the results of section 2. 
To achieve this it is convenient 
to start with the differential statement
\eqlabel{\differentialrln}
$$d\G^{(2)}_t=\G^{(2,1)}_t dt+{1\over6}\G^{(4)}_td\bar\varphi^2\no$$
If we integrate this relation along a contour of constant magnetization, $\vfb$,
from the critical isotherm $t=0$ we obtain
\eqlabel{\integ}
$$\G^{(2)}_t(t,\vfb)=\G^{(2)}_t(0,\vfb)+\int_0^t{\G^{(2,1)}_t(t',\vfb)}dt' \no$$
{}From the definitions of the Wilson functions implied by the normalization 
conditions (\docref{ncone}--\docref{ncfour}) on inverting the relation
(\docref{integ}) we find
\eqlabel{\eqofstafin}
$$t+\int^{\ss{M(0,\vfb)}}_0{xdx}(2-\gf(x))
e^{-\int_\kappa^x\gft(x'){dx'\o x'}}
=\int_0^{\ss{M(t,\vfb)}}{x dx}(2-\gf(x))
e^{-\int_\kappa^x\gft(x'){dx'\o x'}}\no$$

Above we have the relation $M=M(t,\vfb)$, but parametrically in 
terms of $M(0,\vfb)$. We
can determine the latter as an explicit function of $\vfb$ by integrating 
along the critical isotherm from the critical point. 
If we integrate (\docref{differentialrln}) along the critical isotherm 
up from the critical point we find
\eqlabel{\gtt}
$$\G^{(2)}_{t}(0,\vfb)=\int_0^{\vfb^2}{\G_{t}^{(4)}(x)\over6}dx\no$$
By choosing the normalization point $NP$ in 
(\docref{ncone}--\docref{ncfour})  on the critical isotherm such that 
the transverse correlation length is $\kappa^{-1}$ and at zero momentum, we 
can express $\G^{(4)}_t$ in the form 
$$\G^{(4)}_t=\l(\kappa)\ {\rm exp}\int_\kappa^{M(0,\vfb)}(\tilde\gl(x)-2\tilde\gft(x))\no$$ 
where $\tilde\gl(x)$ and $\tilde\gft(x)$ are the resulting  Wilson functions 
from this prescription.
Inverting (\docref{gtt}) one finds
\eqlabel{\phirel}
$$\vfb^2={6\o\l}\int_0^{M(0,\vfb)}(2-\tilde\gf(x))
e^{-\int_\kappa^x(\tilde\gl(x')-\tilde\gf(x')){dx'\o x'}}{x dx}\no$$

The two equations (\docref{eqofstafin}) and 
(\docref{phirel}) specify completely the relation between the 
transverse correlation length, temperature and magnetization.
Finally, since the transverse correlation length is infinite on the co-existence 
curve, i.e. $M(t_{coex},\vfb)=0$ the equation of the co-existence curve is given by
\eqlabel{\eqofcoex}
$$t+\int^{\ss{M(0,\vfb)}}_0{x dx}(2-\gf(x))
e^{-\int_\kappa^x\gft(x'){dx'\o x'}}
=0\no$$
where $M(0,\vfb)$ as a function of $\vfb$ is determined by 
equation (\docref{phirel}). 

The specific heat and the energy density are 
proportional to $\G^{(0,2)}$ and $\G^{(0,1)}$ respectively. We can treat these in 
a similar fashion to the above by beginning with the differential relation
$$d\G^{(0,2)}=\G^{(0,3)}dt+{1\over2}\G^{(2,2)}_td\vfb^2\no$$
By integrating along a contour of constant $\vfb$ 
up from the co-existence curve we obtain
$$\G^{(0,2)}=\int_0^M(2-\gf(x))e^{2\int^x_\kappa\gft(x)}\bar\G^{(0,3)}(x)x^{d-5}dx\no$$
and
$$\G^{(0,1)}=\int^M_0(2-\gf(x))e^{-\int^x_\kappa\gft(x)}xdx
\int_0^x(2-\gf(x'))e^{2\int^{x'}_\kappa\gft(x')}\bar\G^{(0,3)}(x'){x'}^{d-5}dx'\no$$
where 
$$\bar\G^{(0,3)}(M)={\G^{(0,3)}{(\G^{(2)})}^3\over {(\G^{(2,1)})}^3M^d}\no$$
The advantage of integrating up from the co-existence curve is that
we can extract the singular part by requiring that
both $\G^{(0,1)}$ and $\G^{(0,2)}$ vanish there.
We can impose such boundary conditions only at the critical point
in the special case of when $\alpha<0$ which is the case for the large $N$
limit.  A boundary condition at some other point is also possible but the 
formulae are more complicated.

For the large $N$ limit we can evaluate the function 
$\bar\G^{(0,3)}$ exactly. We find diagrammatically that
$$\bar\G^{(0,3)}=N{\itro\over M^{d-6}}\no$$
which can be related to derivatives of the scaling function
${\cal F}$. It is now a simple matter of performing the relevant 
integrations to recover the renormalization group results for these
functions. 

\subsec{Equation of State}
In the large $N$-limit one loop RG expressions 
become exact. Diagrammatically the renormalization constants 
$Z_{\vf}$, $Z_{\ft}$ and $Z_{\l}$ are
\eqlabel{\reconst}
$$Z_{\ft}=Z_{\l}=1-{\l_BN\o6}\bub,
\ \ \ \ \ \ \ \ \ \ \ \ \ \ \ \ Z_{\vf}=0\no$$
Note that $\bub$ is an explicit function of $z=\k L$, with $\k$ the 
fiducial finite size correlation length used as a running 
parameter.

In terms of the floating coupling [\cite{NPhysOrig}] $h$, chosen so as to make the 
coefficient of the quadratic term in the resulting $\beta$ function unity, one finds 
\eqlabel{\betafn} 
$$\beta(h,z)=-\e(z) h+h^2\no$$ 
and
\eqlabel{\gftfn}
$$\gft(h,z)=\gl(h,z)=h, \qquad\qquad\gf=0\no$$
where the function $\e(z)$ is
$$\e(z)={6z^2\bfour(z)\over\itro(z)}-2.\no$$  
Similarly for the critical isotherm we have
\eqlabel{\gftfntilde}
$$\tilde\gft(h,z)=\tilde\gl(h,z)=h, \qquad\qquad\tilde\gf=0\no$$

Equations (\docref{gftfn}) and (\docref{gftfntilde})  implies that $\tilde\gl=\gft$ 
and hence with (\docref{phirel}) we find 
(\docref{eqofstafin}) becomes 
\eqlabel{\rgconstraint}
$$t_0+{\l_0\vfb^2\o6}
={2\over L^2}\int_0^{z}{dx\o x}
e^{\int_{z_0}^x(2-h(x')){dx'\o x'}}\no$$
where $h$, the solution of (\docref{betafn}), is 
\eqlabel{\betahNol}
$$h(z,z_0,h_0)={e^{-\int_{z_0}^{z}\e(x){dx\o x}}\o{h_0^{-1}
-\int_{z_0}^{z}e^{-\int^x_{z_0}\e(x'){dx'\o x'}}{dx\o x}}}\no$$ 
with $z_0=\kappa_0 L$, $\l_0=\l(\kappa_0)$ and $t_0=t(M,\kappa_0)$.
The steepest descent constraint of the large $N$ limit 
has now been recovered in the form (\docref{rgconstraint}).

In (\docref{betahNol}) the initial coupling is specified  at the
``microscopic'' scale $\k_0$. For $d<4$ this microscopic scale can be 
sent to infinity while maintaining $h_0$ finite.  A universal 
floating coupling,  $h(z)={4z^2\itro(z)\o\bub(z)}$,
which is the separatrix solution of the differential equation is obtained. 
If this solution is used in (\docref{rgconstraint}) we obtain
\eqlabel{\rgeqofstate}
$$t_0+{\l_0\vfb^2\o6}={N\l_0\over 6}
M^{d-2}{\cal F}(d,z)\no$$
where $\l_0={6\over N\bub(\kappa_0)}$ is the initial dimensional
coupling corresponding to the separatrix solution.
In the asymptotic regime (\docref{perbub}) implies 
$\l_0={12\kappa_0^{4-d}\over N(d-2)\sigma_d}$.
We see that we have recovered from our renormalization group arguments
the universal form of the spherical constraint (\docref{invscaling}) where
now $\tau(L)={6t_0\over\l_0}$.
If one is interested in
corrections to scaling, as is usually the case in comparing with 
experimental data, then $\k_0$ should be left finite and fitted to the data.
The cases of $d=3$ and $d=4$ require special care. For $d=4$ one can
not ignore $h_0$ but again when appropriate care is taken one recovers the results
of the previous section.

For periodic boundary conditions 
$$\e(z)=5-d-(7-d){{\displaystyle\sum_{n=-\i}^{\i}{4\pi^2n^2\over z^2}
\left(1+{4\pi^2n^2\over z^2}\right)^{d-9\over2}}
\over{\displaystyle\sum_{n=-\i}^{\i}\left(1+{4\pi^2n^2\over
z^2}\right)^{d-7\over2}}}\no$$ 
and the separatrix coupling
\eqlabel{\perh}
$$h(z)=(5-d){{\displaystyle\sum_{n=-\infty}^{\infty}}{(1+{({2\pi n\o
z})}^2)}^{(d-7)\o2}
\o{\displaystyle\sum_{n=-\infty}^{\infty}}{(1+{({2\pi n\o
z})}^2)}^{(d-5)\o2}}\no$$
For $d=3$ the results are particularly simple
\eqlabel{\htd} $$h(z)=1+{z\o\sinh z}\no$$
\eqlabel{\epstd}
$$\e(z)=1+{z^2\coth({z\o2})\o\sinh z+z}\no$$
With the explicit results above one recovers (\docref{fscaln}) for ${\cal F}$ 
and hence ${\cal Q}$.

\beginsection{Effective Exponents in the Large $N$ limit}
We will now put to use the formulae derived in the previous sections
to calculate various scaling functions which describe the crossovers between
$d$ dimensional and $d-1$ dimensional critical behavior, and the crossover 
to mean field theory. A very useful way of representing a large class of scaling functions 
is in terms of effective
critical exponents, which are functions that interpolate 
between the constant critical exponents associated with the different asymptotic 
regimes that characterize the crossover. 

First of all we define, for $T>T_c(L)$ and $H=0$, an effective critical 
exponent $\nu\ef$ 
\eqlabel{\nueffdef}
$$\nu\ef=-\left.{d\ln\xi_L\o d\ln t_B}\right\vert_{\ss H=0}\no$$
where $\xi_L$ is the correlation length associated with the transverse 
dimensions, i.e. the correlation length in the infinite dimensions 
(remember that $t_B\sim T-T_c(L)$). 
As ${1\o\xi_L}=M$  one finds from  (\docref{msquare}) that
\eqlabel{\nueffdiag}
$$\nu\ef={1\o2}\left({1-{\l_BN\o6M^2}{\io'}\o{1+{\l_BN\o6}\bub}}\right)\no$$
Explicitly
\eqlabel{\nueffmf}
$$\nu\ef=\left({1+{6\o g{\cal F}}\o d-2+{d\ln {\cal F}\o d\ln z}+
{12\o g{\cal F}}}\right)\no$$
where $z=ML$ and the scaling function ${\cal F}$, for periodic 
boundary conditions is given by (\docref{fscaln}).

As the coupling constant $\l_B$ is present in the scaling function it is not universal
in the normal sense of the word. The coupling $g$ governs the corrections to
scaling about the critical theory. In the limit $g\ra0$ one crosses over to mean field 
theory where
$\nu\ef\ra{1\o2}$. In the critical regime, where $z\ll g$ and $1\ll g$,
the terms proportional to $g^{-1}$ may be neglected yielding
\eqlabel{\nueffuni}
$$\nu\ef=\left(d-2+{d\ln{\cal F}\o d\ln z}\right)^{-1}\no$$
which is now a true universal scaling function. We can also think of getting rid of the
corrections to scaling by keeping a cutoff $\L$ in the diagrams and
making an appropriate choice of the non-universal coupling $\l_B$. 
To see this we divide the integration range 
$[0,\L]$ into two parts $[0,\i]$ and $[\i,\L]$. In the limit $\L L\gg1$ the
integrand in the 
integration over the range $[\i,\L]$ can be approximated by the ``bulk''
expression to yield $M^2{N\l_B\o3(4\pi)^{d\o2}\G({d\o2})}{\L^{d-4}\o(d-4)}$ up to 
corrections $O({\rm exp}(-\L L))$. Hence, the choice $\l_B={3(4-d)\o N}
(4\pi)^{d\o2}\G({d\o2})\L^{4-d}$ will eliminate the corrections to scaling. 
{}From (\docref{nueffuni}) we see that as $z\ra0$ then $\nu\ef\ra{1\o(d-3)}$ 
whereas for $z\ra\i$, $\nu\ef\ra{1\o(d-2)}$. Thus $\nu\ef$ interpolates between 
the two exact asymptotic values associated with the spherical model in $d$ and 
$d-1$ dimensions.
 
We can also consider an effective exponent 
$\gamma\ef=-{d\ln\chi\o d\ln t_B}$ where $\chi$ is the susceptibility for $H=0$. 
As trivially $\chi=M^{-2}$, one obtains
\eqlabel{\gammaeffmf}
$$\gamma\ef=2\left({1+{6\o g{\cal F}}\o d-2+{d\ln {\cal F}\o d\ln z}+
{12\o g{\cal F}}}\right)\no$$
In the mean field limit $\gamma\ef\ra1$ whereas we have
$\gamma\ef\ra{2\o(d-3)}$ as $z\ra0$ and 
$\gamma\ef\ra{2\o(d-2)}$  as $z\ra\i$. Thus once again we see the effective
exponent interpolates between the known asymptotic fixed points of the model.

If we consider what happens for $T<T_c(L)$ and $H=0$, the effective
exponents corresponding to $\nu\ef$ and $\gamma\ef$ are 
ill defined, however, the effective exponent 
$\beta\ef={d\ln\vfb\o d\ln\vert t_B\vert}$ is well defined. 
{}From the saddle point equation
(\docref{consttwo}), as ${\io'}$ vanishes, 
due to the vanishing of the transverse mass
on the coexistence curve,  we see
from (\docref{rgeqofstate}) that 
$$\vfb^2 ={6t\over \l}\no$$
which implies that $\beta\ef=1/2$, i.e. there is no
crossover as one proceeds along the coexistence curve. This is in strong
distinction to the Ising model where there is a crossover between the critical point and
the strong coupling discontinuity fixed point at $T=0$. The
contrast is due to the fact that the coexistence curve is a line of first order transitions
for $N=1$ and a line of continuous transitions for $N>1$.

Considering now the approach to the critical point as a function of field, $H$, 
on the critical isotherm, $T=T_c(L)$,
we define an effective exponent $\delta\ef=\left.{d\ln H\o d\ln\vfb}\right\vert_{\ss t_B=0}$.
This implies that 
$$\delta\ef = 1+{\G^{(4)}_t\vfb^2\over 3\G^{(2)}_t}\no$$
{}From (\docref{constone}) at $T=T_c(L)$ one finds
\eqlabel{\deltadiag}
$$\delta\ef=\left({3+{\l_BN\o6}\bub-{\l_BN\o3M^2}{\io'}\o{1+{\l_BN\o6}\bub}}\right)\no$$
or in terms of the universal scaling function ${\cal F}$
\eqlabel{\deltamf}
$$\delta\ef=\left({d+2+{d\ln{\cal F}\o d\ln z}+{36\o g{\cal F}}
\o{d-2+{d\ln{\cal F}\o d\ln z}+{12\o g{\cal F}}}}\right)\no$$
In the mean field limit $\delta\ef\ra3$ whilst in the universal limit, when we eliminate
the corrections to scaling, we find
\eqlabel{\deltauni}
$$\delta\ef=\left({d+2+{d\ln{\cal F}\o d\ln z}
\o{d-2+{d\ln{\cal F}\o d\ln z}}}\right)\no$$

The specific heat and the energy density may also be discussed 
in terms of effective exponents. 
We may define, $\alpha^s\ef=-{d\ln {\cal C}\o d\ln t_B}$ and 
$1-\alpha^e\ef={d\ln{\cal E}\o d\ln t_B}$, where ${\cal C}$ is the specific heat  
and ${\cal E}$ is the energy density for $H=0$ respectively. 
{}From the definition of $\alpha^e\ef$ we have
$$\alpha^e\ef=1-{t_B\G_B^{(0,2)}\over\G_B^{(0,1)}}\no$$
and using (\docref{gzeroone}) and (\docref{gzerotwoba}) we get diagrammatically
$$\alpha^e\ef={\l_BN\o6}\left({\bub+{\io'\o M^2}\o 1+{\l_BN\o6}\bub}\right)\no$$
hence
\eqlabel{\alphaeeffmf}
$$\alpha^e\ef=\left({d-4+{d\ln{\cal F}\o d\ln z}\o
{d-2+{d\ln{\cal F}\o d\ln z}+{12\o g{\cal F}}}}\right)\no$$
In the mean field limit $\alpha\ef\ra0$ as expected, whilst eliminating the 
corrections to scaling yields the universal function
\eqlabel{\alphaeeffuni}
$$\alpha^e\ef=\left({d-4+{d\ln{\cal F}\o d\ln z}\o
{d-2+{d\ln{\cal F}\o d\ln z}}}\right)\no$$

Turning to the specific heat, diagrammatically
$$\alpha^s\ef=-{\l_BN\o3}M^{2}\itro{(1-{\l_BN\o6}\io ')\o (1+{\l_BN\o6}\bub)^2}\no$$
and in terms of the scaling function ${\cal F}$
\eqlabel{\alphaseffmf}
$$\alpha^s\ef=\left(1+{6\o g\cf}\right)
{((d-2)(d-4)+4(d-2){d\ln{\cal F}\o d\ln z}
+{4z^4\o{\cal F}}{d^2{\cal F}\o d(z^2)^2})\o 
(d-2+{d\ln{\cal F}\o d\ln z}+{12\o g{\cal F}})^2}\no$$
This also vanishes in the mean field limit. The corresponding universal scaling 
function without corrections to scaling is
\eqlabel{\alphaseffuni}
$$\alpha^s\ef={((d-2)(d-4)+4(d-2){d\ln{\cal F}\o d\ln z}
+{4z^4\o{\cal F}}{d^2{\cal F}\o d(z^2)^2})\o (d-2+{d\ln{\cal F}\o d\ln z})^2}\no$$
It is easy to verify that the two scaling functions interpolate between exactly the same
asymptotic limits and that in particular the universal ones
interpolate between $\alpha(d')$ and $\alpha(d)$ as $z$ ranges from zero 
to infinity. For $T<T_c$ on the coexistence curve the 
singular parts of the energy density and the specific heat are identically zero and
thus either $\alpha^e\ef=\alpha^s\ef=0$ or the amplitudes associated 
with these exponents are zero in this regime. In the current model it is the latter
that is in fact the case.

\beginsection{\bf  Effective Exponent Scaling Laws\hfill}
In this section we wish to make some observations concerning certain algebraic relations
between the effective exponents. First of all from equations 
(\docref{nueffmf}), 
(\docref{gammaeffmf}) and the fact that $\eta\ef\equiv0$ due to the 
vanishing of $\gf$ in the large $N$ limit, 
we see that the following relation is valid
\eqlabel{\scalawone}
$$\gamma\ef=\nu\ef(2-\eta\ef)\no$$
Note that this is true even for the scaling functions that include the crossover to mean
field theory as well as the universal crossover between the $d$ and $d-1$-dimensional 
critical points. {}From the expressions for $\gamma\ef$, $\a{\ef}$ 
and $\beta\ef$ we see that
\eqlabel{\scalingrelation}
$$\a^e\ef+2\beta\ef+\gamma\ef=2\no$$
also in the general case.
The exponent $\a^s\ef$ however, does not satisfy this relation. We see then that
direct analogs of the normal scaling laws between exponents hold, 
however, here the relations are between entire
scaling functions

It is natural to ask if there are analogs of other scaling laws, in particular hyperscaling 
where $2-\alpha=\nu d$. {}From equations (\docref{nueffmf}), 
(\docref{alphaeeffmf}) and
(\docref{alphaseffmf}) it is clear that for fixed dimensionality $d$, $2-\alpha^i\ef\neq\nu\ef d$
where $i=e,s$. However, one can define the notion of an effective dimensionality $d^i\ef$
such that an effective hyperscaling law is valid. Defining $d^e\ef=(2-\alpha^e\ef)/\nu\ef$ 
one finds
\eqlabel{\deffedef}
$$d^e\ef=\left({d+{d\ln{\cal F}\o d\ln z}+{24\o g{\cal F}}\o 1+{6\o g{\cal F}}}\right)\no$$
In the mean field limit $d^e\ef\ra4$, i.e. the upper critical dimension as one might expect.
In the limit $z\rightarrow\infty$, one finds $d^e\ef\rightarrow d$, whilst in the 
limit $z\rightarrow0$
for fixed $L$,  $d^e\ef\ra(d-1)$. Of course one could also define an effective 
dimensionality via $\alpha^s\ef$ as $d^s\ef=(2-\alpha^s\ef)/\nu\ef$. 
The expression is unwieldy so we will not
write it down. Again it interpolates between 
$4$, $d$ and $d-1$ in the appropriate asymptotic limits. However since
the exponent $\alpha^s\ef$ did not satisfy the scaling law 
(\docref{scalingrelation}) we do not expect $d^s\ef$ to satisfy corresponding
laws and in fact it doesn't.

We might enquire now as to the validity  of other scaling laws that involve the dimensionality
explicitly. Noticing that $d^e\ef=2+{1\o\nu\ef}$ (i.e. $\nu\ef=1/(d^e\ef-2)$)
one arrives at the scaling relation
\eqlabel{\scalawtwo}
$$\beta\ef={\nu\ef\o2}(d^e\ef-2+\eta\ef)\no$$
In like fashion one finds that
\eqlabel{\scalawthree}
$$\delta\ef=\left({d^e\ef+2-\eta\ef\o d^e\ef-2+\eta\ef}\right)\no$$

  
We conclude here then that the effective exponents for the $N\ra\i$ limit of an $O(N)$
model satisfy natural analogs of the normal scaling laws, including hyperscaling,  
if we introduce the notion of an effective dimension $d^e\ef$. What are we to make of this? 
In  the standard case of a single critical point the scaling laws tell us that out of all 
the critical exponents, which remember are numbers, only two are independent, 
hence a specification of two exponents, 
and the dimensionality of the system to account for hyperscaling, is sufficient.
In the limit $N\ra\i$ we know that knowledge of only one exponent 
is necessary due to the triviality of $\eta$. {}From a field theoretic RG 
point of view we
can rephrase this by saying that only one renormalization constant is needed; to
renormalize the operator $\vf^2$, no 
renormalization of $\vf$ is required. In the crossover case, where
one has to implement a global, non-linear RG as opposed to simply linearizing around
a given fixed point, we are being told that of the six scaling functions
$\alpha^e\ef$, $\nu\ef$, $\gamma\ef$, $\eta\ef$, $\delta\ef$ and $\beta\ef$
only one is independent, and that all relevant information can be encoded in the one
function $d^e\ef$. 

\beginsection{\bf Conclusions\hfill} 

In this paper we studied dimensional crossover for a $d$ dimensional film
with periodic boundary conditions in an exactly solvable model --- the
large $N$ limit of an $O(N)$ Landau-Ginzburg-Wilson model, a model that
also includes crossover to mean field behavior.  We obtained the scaling
forms of the free energy and equation of state and extracted their
universal limits where the crossover to mean field theory is eliminated.
We studied the model using both direct methods and the techniques of
environmentally friendly renormalization and were therefore able to
compare this RG approach with an exact solution obtained by other means. 

In the RG approach the equation of state was found by choosing
normalization conditions at a fiducial value of the transverse correlation
length and integrating it along particular contours in the $(t,\vfb)$
phase diagram; first along a contour of constant magnetization and second
along the critical isotherm. The first was equivalent to a coordinate
change $(t,\vfb)\ra(M,\vfb)$. Both contours together yielded a complete
description of the phase diagram of the model. 

A natural set of effective exponents were defined and computed
which also exhibited both dimensional crossover and the crossover to mean
field exponents. By taking the universal limit we obtained effective
exponents for the universal dimensional crossover. Finally we noticed
that the effective exponents in the model obeyed natural analogs of the
standard scaling laws. In the case of hyperscaling this was only true if
one defined an effective dimensionality $\de$ which interpolated between
the two asymptotic limits $d$ and $d-1$. In this particular model it was
found that all the effective exponents could be written as functions of
$\de$ and hence that there was really only one independent scaling
function. 

A natural question is whether or not the effective exponent laws extend to
the case $N\neq\i$. The validity of effective exponent laws was
discussed in the wider context of environmentally friendly renormalization
in [\cite{EnvRG}], and using RG arguments it was shown there that certain
scaling relations between thermodynamic functions exist which are a
natural generalization of the scaling laws to crossover problems. 

{}From a theoretical point of view the underlying object governing all
effective exponents, whether derived from the energy ${\cal E}$ or the
specific heat ${\cal C}$, is the singular part of the free energy, 
$k_BT\tilde\G$. Other thermodynamic quantities are derived from $\tilde\G$ by
differentiation. Singular ones then have two asymptotic scaling forms, for
example a thermodynamic function $P$ has the form $P\sim
A_{d'}^{\pm}(L)\vert T-T_c(L)\vert^{-{\theta_{d'}}}+S_d L^{-\varphi(d)}$
in the neighborhood\footnote{$^*$}{\tenpoint Generally for continuous
transitions one would expect $S_d^{+}=S_d^{-}$ and we assume this here.}
of $T_c(L)$, and for $L\ra\i$, $T\ra T_c(\i)$ it takes the form $P\sim
A_d^{\pm}\vert T-T_c(\i)\vert^{-{\theta_{d}}}$.  Quite generally one can
decompose the scaling function $P$ throughout the crossover into the form
\eqlabel{\gendecomp} 
$$P=A^{\pm}(t,L)\exp^{-\int_1^t\theta(x){dx\o x}}+S(t,L)\no$$ 
where at the respective asymptotic end points $A^{\pm}$ gives the
amplitude, $\theta$ the exponent and $S^{\pm}$ any shift that may have
arisen in the thermodynamic function. However, in the crossover region
this division is somewhat arbitrary. In the case of an effective exponent
defined as the logarithmic derivative of $P_s=P-S_d/L^{\varphi(d)}$ with
respect to $T-T_c(L)$ the decomposition is forced to take the form
\eqlabel{\effdecomp}
$$P=A_d^{\pm}(L)\exp^{-\int_1^t\theta\ef(x){dx\over x}}
+S_d L^{-\varphi(d)}\no$$ 
In the large $N$ limit we have found that the above decomposition
has the further property that for an 
appropriate definition of effective exponents, all the usual scaling 
relations are obeyed. 

We see that the basic reason such scaling laws are obeyed in the large $N$
limit is because the basic building function $\tilde\Gamma$ involves $t$
and $\vfb$ only in the combination
$$w=(\tilde\varphi^{2}+\tau) L^{1/\nu}+b_d$$ 
so that derivatives with respect to $\tau$ and with respect to $\varphi$
are intimately related. This explains why the energy rather than the
specific heat provided the effective exponent that yielded the scaling
laws.  It is difficult to expect such a reduction to one scaling variable
for more general models. Rather the two variables $\tilde x$ and 
$\tilde y$ should enter separately. 

In general the decomposition (\docref{effdecomp}) may not be possible, and
is not expected to give effective exponents that obey scaling laws. 
However, we can choose a rather natural division into amplitude and
floating effective exponent in the form (\docref{gendecomp}) where the
floating effective exponents do obey all the usual scaling relations
including hyperscaling.  A particularly convenient choice is that
associated with separatrix exponents as advocated in [\cite{Btcfss}] where
the basic building functions for the exponents were $\gf$, $\gft$ and
$\gl$ evaluated on the separatrix solution of the RG flow that connects
the $d$ and $d'$ fixed points. Under such a division the amplitudes are
non-singular functions of $t$ and $L$ that interpolate between the $d$ and
$d'$-dimensional amplitudes and the exponents capture all the singular
behavior in the scaling functions. The shift is unaffected by this choice
and retains the form $S_dL^{-\varphi(d)}$. 

\noindent{\bf ACKNOWLEDGEMENTS}

This work was supported in part by Conacyt under grant number
211085-5-0118PE.
\vskip .5truecm

\centerline{\bf REFERENCES}

\item{[\cite{NPhysOrig}]}Denjoe. O'Connor and C. R. Stephens, 
{\it Nucl. Phys.} {\bf B360} (1991) 297; {\it J.Phys.} {\bf A25}
(1992) 101.

\item{[\cite{Btcfss}]}F. Freire, Denjoe O'Connor and C.R. Stephens, 
{\it J. Stat. Phys.} {\bf 74} (1994) 219.

\item{[\cite{proc}]} C.R. Stephens, {\it Jou. Mag. Mag. Mat.} 
{\bf 104-107} (1992) 297; Denjoe O'Connor and C. R. Stephens, 
{\it Proc. Roy. Soc.} {\bf 444} (1994) 287.

\item{[\cite{fintemp}]} Denjoe O'Connor, C.R. Stephens and F. Freire,  
{\it Mod. Phys. Lett.} {\bf A8} (1993) 1779; 
M.A. van Eijck, Denjoe O'Connor and C.R. Stephens, {\it Int. J. Mod. Phys.} 
{\bf A23} (1995) 3343.

\item{[\cite{prl}]} Denjoe O'Connor and C. R. Stephens, 
{\it Phys. Rev. Lett.} {\bf 72} (1994) 506.
 
\item{[\cite{EnvRG}]}Denjoe O'Connor and C.R. Stephens, 
{\it Int. J. Mod. Phys.} {\bf A9} (1994) 2805.

\item{[\cite{baker}]}G.A. Baker, B.G. Nickel and D.I. Meiron, 
{\it Phys. Rev.} {\bf B17} (1978) 1365.

\item{[\cite{binder}]}K. Binder, {\it Phase Transitions and Critical
Phenomena} Vol. {\bf 5B}, edited by Domb and Green (1976). 

\item{[\cite{capefish}]}T.W. Capehart and M.E. Fisher, 
{\it Phys. Rev.} {\bf B13}, (1976) 5021.

\item{[\cite{BerlinKac}]}T.H. Berlin and M. Kac, {\it Phys. Rev.} {\bf 89}
(1952) 821. 

\item{[\cite{Stanley}]}H.E. Stanley, {\it Phys. Rev.} {\bf 176} (1968) 718.

\item{[\cite{Wilson}]}K.G. Wilson, {\it Phys. Rev.} {\bf D7} (1973) 2911.

\item{[\cite{BreWadia}]}{\it The Large $N$ Limit in Quantum Field Theory
and Statistical Physics} edited by 
E. Br\'ezin and S.R. Wadia, World Scientific 1993.

\item{[\cite{BarberFisher}]} M. Barber and M.E. Fisher, {\it Ann. Phys.}, {\bf
77} (1973) 1. 

\item{[\cite{Rudnick}]}J. Rudnick, {\it Finite Size Scaling and 
Numerical Simulations of Statistical Systems} edited by V. Privman, 
World Scientific 1990.

\item{[\cite{AllenPathria}]} S. Allen and R.K. Pathria, {\it Phys. Rev.} {\bf B50}
(1994) 6765.

\item{[\cite{spheat}]} F. Freire, Denjoe O'Connor and C.R. Stephens,
{\it Phys. Rev.} {\bf E53} (1996) ??. 

\item{[\cite{Joyce}]} G.S. Joyce, {\it Phase Transitions and Critical Phenomena}
Vol. {\bf 2} edited by C. Domb and M.S. Green, Academic Press 1972.

\bye